\documentclass[%
aip,
jmp,%
amsmath,amssymb,
reprint,%
]{revtex4-1}

\usepackage{graphicx}
\usepackage{dcolumn}
\usepackage{bm}
\usepackage{color}
\usepackage{CJK}
\usepackage{fancyhdr}
\usepackage{calc}
\usepackage{float}
\usepackage{mpgmpar} 
\usepackage{amsmath,amsthm} 
\usepackage{bm,here}
\usepackage{lipsum}
\usepackage{booktabs}

\begin{document}

\preprint{AIP/123-QED}
\title[]{Ultrafast dynamics of electronic structure in InN thin film}

\author{Junjun Jia}
\email{jia@aoni.waseda.jp}
\affiliation{%
Global Center for Science and Engineering (GCSE), Faculty of Science and Engineering, Waseda University, 3–4–1 Okubo, Shinjyu-ku, Tokyo 169--8555, Japan.
}%

\author{Takashi Yagi}
\affiliation{%
National Metrology Institute of Japan (NMIJ), National Institute of Advanced Industrial Science and Technology (AIST), Central 5, 1-1-1 Higashi, Tsukuba, Ibaraki 305-8565, Japan
}%

\author{Toshiki Makimoto}
\affiliation{%
Graduate School of Advanced Science and Engineering, Waseda University, 3–4–1 Okubo, Shinjyu-ku, Tokyo 169--8555, Japan.
}%


\date{\today}
\begin{abstract}
Simultaneous measurements of transient transmission and reflectivity were performed in the unintentionally doped InN film to reveal ultrafast optical bleaching and its recovery behavior under intense laser irradiation. The optical bleaching is attributed to Pauli blocking due to the occupation of photoexcited electrons at the probing energy level. The time constant for the transition from the excitation state to the conduction band edge is $\sim$260 fs. The interplay between band filling and band gap renormalization caused by electron--hole and electron--electron interactions gives rise to complex spectral characteristics of transient reflectivity, from which the time constants of photoexcited electron--hole direct recombination and band edge recombination are extracted as $\sim$60 fs and 250$\sim$400 fs, respectively. Our results also reveal that the electron--electron interaction suppresses band edge recombination, and mitigates the recovery process. Our experiments highlight the controllability of the band structure of semiconductors by intense laser irradiation.  

\end{abstract}

\maketitle
\section{Introduction}
InN thin films have attracted considerable attention due to their potential application as a saturable absorber or as a component in ultrafast all-optical switching devices, because of large optical bleaching effect and ultrafast bleaching recovery.\cite{Pacebutas2006, Ricardo2006, Tsai2007, Sun2008, Ahn2009, Ahn2012} The optical transmittance of these films increases nearly 5 times under intense laser irradiation, and recovery to the original state occurs on the order of picoseconds.\cite{Pacebutas2006} Such striking optical characteristics highlight the need for a comprehensive understanding of photoexcited carrier (electron and hole) dynamics to elucidate the mechanisms involved. 

Thus far, all reported techniques for revealing the ultrafast carrier dynamics during optical bleaching are based on pump--probe measurements, {\it  i.e.}, transient transmission or transient reflectivity. Optical bleaching in InN was first discovered by the traditional Z-scan technique and transient transmission measurements.\cite{Pacebutas2006} This phenomenon is generally postulated to originate from band filling (Burstein--Moss effect) due to the injection of photoexcited electrons into the conduction band (CB),\cite{Pacebutas2006, Ahn2012} when the photon energy is larger than the narrow direct band gap (0.7 eV) of InN.\cite{Wu2002, Yu2005} In order to obtain more understanding about such abnormal optical bleaching, most studies have applied the pump--probe transient reflectivity measurements to investigate ultrafast carrier dynamics during the recovery process after excitation. The time constants of bleaching recovery range from subpicoseconds to a few hundred picoseconds, and various carrier scattering mechanisms have been proposed to explain such ultrafast recovery behavior, such as the hot phonon effect,\cite{Pacebutas2006} polar optical phonon scattering,\cite{Sun2008} electron--electron scattering,\cite{Tsai2007}, Auger recombination,\cite{McAllister2018}, and nonradiative recombination,\cite{Ricardo2006} as well as the influence of drift and diffusion velocity of carriers on carrier relaxation.\cite{Ahn2009, Ahn2012} However, such carrier scattering mechanisms may not correspond to optical bleaching. Particularly, when the photon energy of the pump laser is much higher than the band gap of InN and there are not enough photoexcited electrons to fill the energy bands up to the probing energy level, the physical phenomena measured by transient reflectivity may differ from those measured by transient transmission. Even though it is clear that the transient reflectivity can reflect ultrafast carrier dynamics, to our knowledge, a comprehensive understanding starting from hot--electron cooling has not yet been sufficiently explored for InN thin films. 

A simple feasible approach for addressing these issues is to simultaneously measure both the transient transmission and transient reflectivity. Capturing both can provide an overall good description of optical bleaching behavior. Surprisingly, thus far, no studies have reported observations of optical bleaching and its recovery behavior in InN thin films via simultaneous measurement, and have provided a comprehensive understanding of the carrier recombination mechanisms and photoexcited carrier lifetimes involved. Undoubtedly, such a comprehensive understanding is important for designing and optimizing InN--based devices.

Herein, we report simultaneous measurements of femtosecond time-resolved transient transmission and reflectivity to investigate ultrafast carrier dynamics in InN thin films. Our experiments clearly revealed three distinct relaxation regimes for photoexcited electrons, specifically, 1) the direct recombination between photoexcited electrons and holes, 2) the relaxation of initial hot carriers, and 3) the recombination of thermalized carriers.

\section{Experimental Details}
\begin{figure}[h]
\begin{center}
\includegraphics[clip, width=8.0cm]{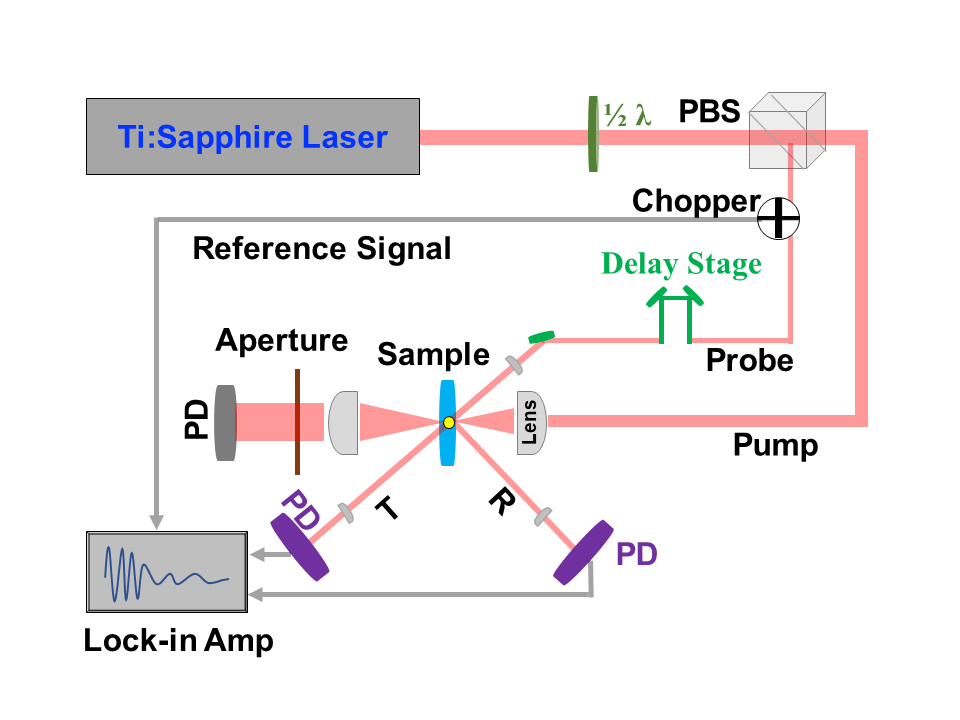}
\caption{Schematic illustration of the experimental setup. A high intensity, horizontally polarized laser pulse with the repetition rate of 76 MHz pumps the InN thin film at normal incidence. The reflection and transmission of a weak probe beam (vertically polarized) are simultaneously recorded. The pump and probe beams are focused by lenses of $f$=4.0 cm and $f$=7.5 cm on the InN sample with a focal spot size less than 10 $\mu$m in diameter. An optical chopper is applied to modulate the probe beam at 205.8 Hz to improve the signal--to--noise ratio. (T: Transmittance, R: Reflectivity, and PD: Photodetector)}
\label{SP}
\end{center}
\end{figure}

Femtosecond time--resolved measurements were performed using an optical pump--probe arrangement, as shown in Fig. \ref{SP}. The unintentionally doped InN thin films were grown on $c$--plane sapphire substrates by RF-assisted molecular beam epitaxy. The carrier concentration and mobility were 6.58$\times$10$^{19}$ cm$^{-3}$ and 28.7 cm$^{2}$/Vs, respectively, which were obtained based on the four--probe method and Hall--effect measurement in the van der Pauw geometry (HL--5500PC, Bio-Rad). The film thickness was 300 nm, as determined by a Jobin Yvon ellipsometer (UVISEL) with an incident and detection angle of 60$^\circ$. 

The InN sample was excited by 140 fs Ti:S laser pulses with a laser pulse repetition rate of 76 MHz at room temperature. The beam was split by a polarized beam splitter. The majority of the power was used to excite the sample (pump), while the remaining power was used as the probe, where the pump and probe laser had the same wavelengths for each measurement. The pump was focused onto the sample surface using a 5$\times$ objective lens (NA=0.14) (Mitutoyo), and the probe was focused on the spot of the pump laser by using an achromatic lens with $f$=7.5 cm. The spatially overlapped spot sizes for the 800--nm pump and probe were less than 10 $\mu$m in diameter, where the pumping and probing incident angles were 90$^\circ$ and 44.8$^\circ$, respectively, with respect to the sample surface. The reflected light was recollimated to enter a Si photodetector (2051--FC--M, New Focus). The probe was chopped at a frequency of 205.8 Hz and lock-in detection was used to improve the signal--to--noise ratio.

\section{Results}
\begin{figure}[h]
\begin{center}
\includegraphics[clip, width=9.0cm]{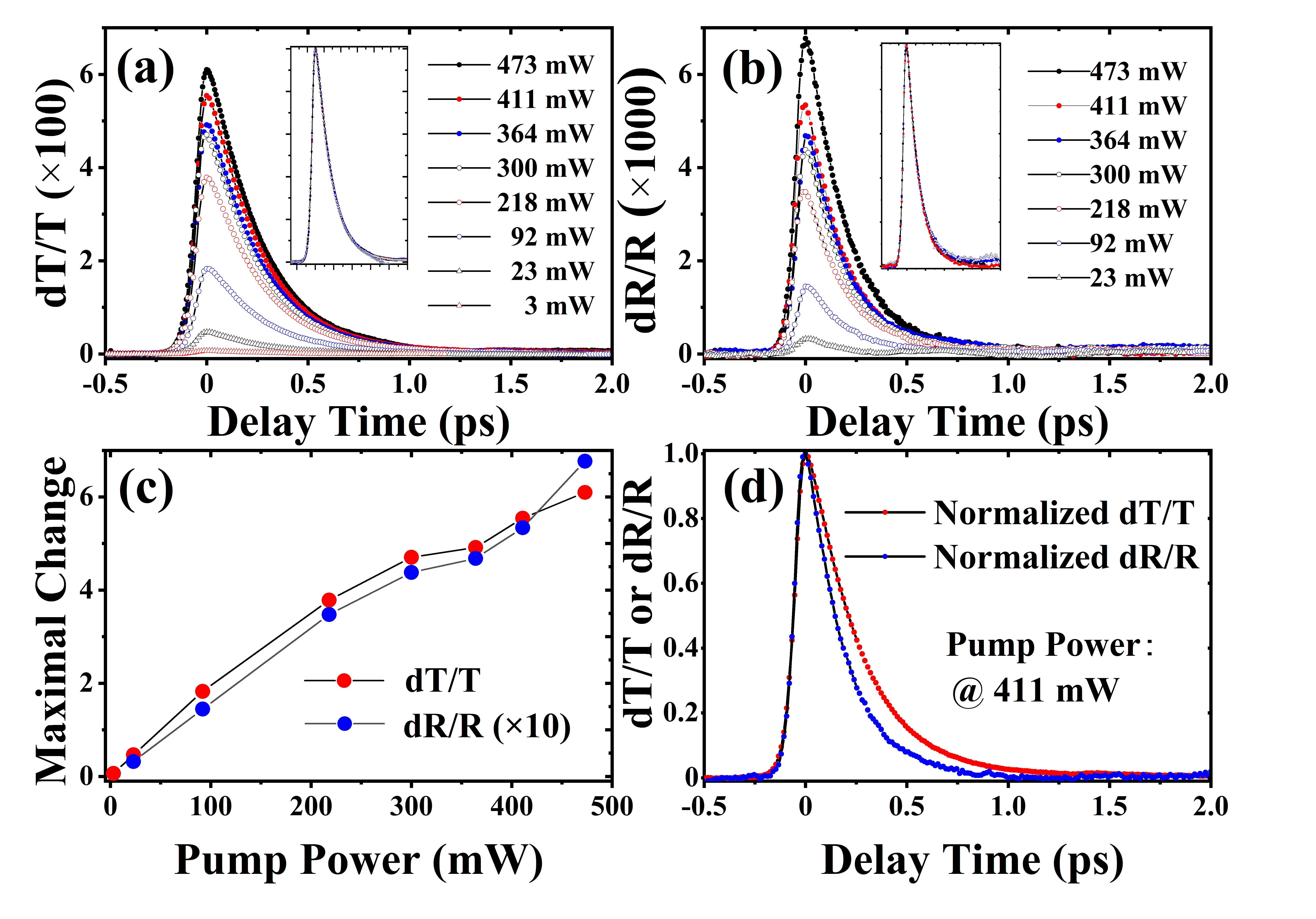}
\caption{Change in transient transmission (a) and reflectivity (b) as a function of time delay, measured at an excitation wavelength of 802 nm. The insets show the normalized transient transmission and reflectivity spectra. (c) Maximal changes in transient transmission and reflectivity spectra at different pump powers. (d) Transient transmission and reflectivity spectra measured at a constant pump power. Here, T and R are the probe transmission and reflectivity, respectively, in the absence of pumping.}
\label{802nm}
\end{center}
\end{figure}

Figures \ref{802nm} (a) and (b) display the measured transient transmission and reflectivity changes at 802 nm for different pump powers. As the pump beam irradiates the sample, the transient transmission and reflectivity of the probe beam rapidly rise, and then decay after reaching a maximum. Interestingly, the change in transient transmission is $\sim$10 fold larger than that for transient reflectivity, as shown in Fig. \ref{802nm} (c). When normalized by the maximum values in Figs. \ref{802nm} (a) and (b), all transmission and reflectivity spectra exhibit the same characteristic shape, as shown in the insets in Fig. \ref{802nm}, suggesting that the recovery behavior is independent of the pump beam power. However, the obvious difference between the spectral characteristics of the transient transmission and reflectivity, shown in Fig. \ref{802nm} (d), implies that these measurements arise from two different physical processes. 

\begin{figure}[h]
\begin{center}
\includegraphics[clip, width=8.0cm]{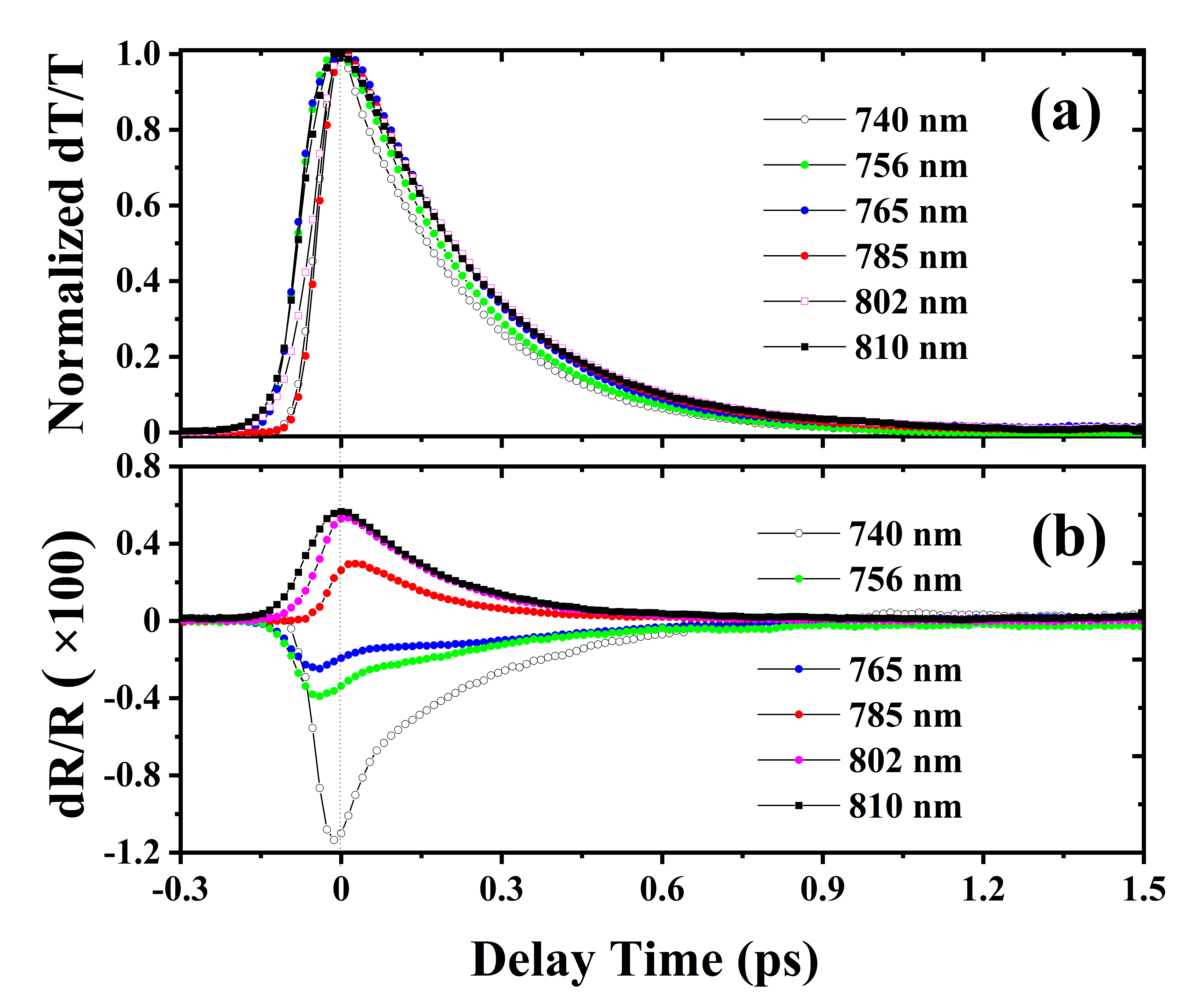}
\caption{Transient transmission and reflectivity of the unintentionally doped InN film with a background carrier density of 6.58$\times$10$^{19}$ cm$^{-3}$ under different excitation wavelengths. All the measurements were performed at a pump power of 411 mW. Each set, where the transient transmission and reflectivity spectra were marked using the same wavelength, was simultaneously measured at the same pumping and probing wavelength, and the origin of the delay time was adjusted to the maximum transient transmission.}
\label{wave}
\end{center}
\end{figure}

To gain more insight into the carrier dynamics during optical bleaching, we investigated the transient transmission and reflectivity at different wavelengths. Fig. \ref{wave} (a) shows the measured transient transmission at excitation photon energies ranging from 1.53 to 1.68 eV (810 to 740 nm) for a fixed pump power of 411 mW. After normalization, all transient transmission curves show the same spectral characteristics. The time constant of the decay curves changes from 260 fs at 810 nm to 226 fs at 740 nm, suggesting that a higher excitation photon energy corresponds to a faster recovery of transient transmission. In contrast, the change in transient reflectivity (dR/R) changes from positive to negative value as the excitation wavelength decreases from 810 to 740 nm. The dR/R is positive above 785 nm, and negative below 785 nm. As the excitation photon energy gradually increases (shorter wavelengths), the absolute magnitude of negative dR/R value increases.    

\begin{figure}[h]
\begin{center}
\includegraphics[clip, width=8.0 cm]{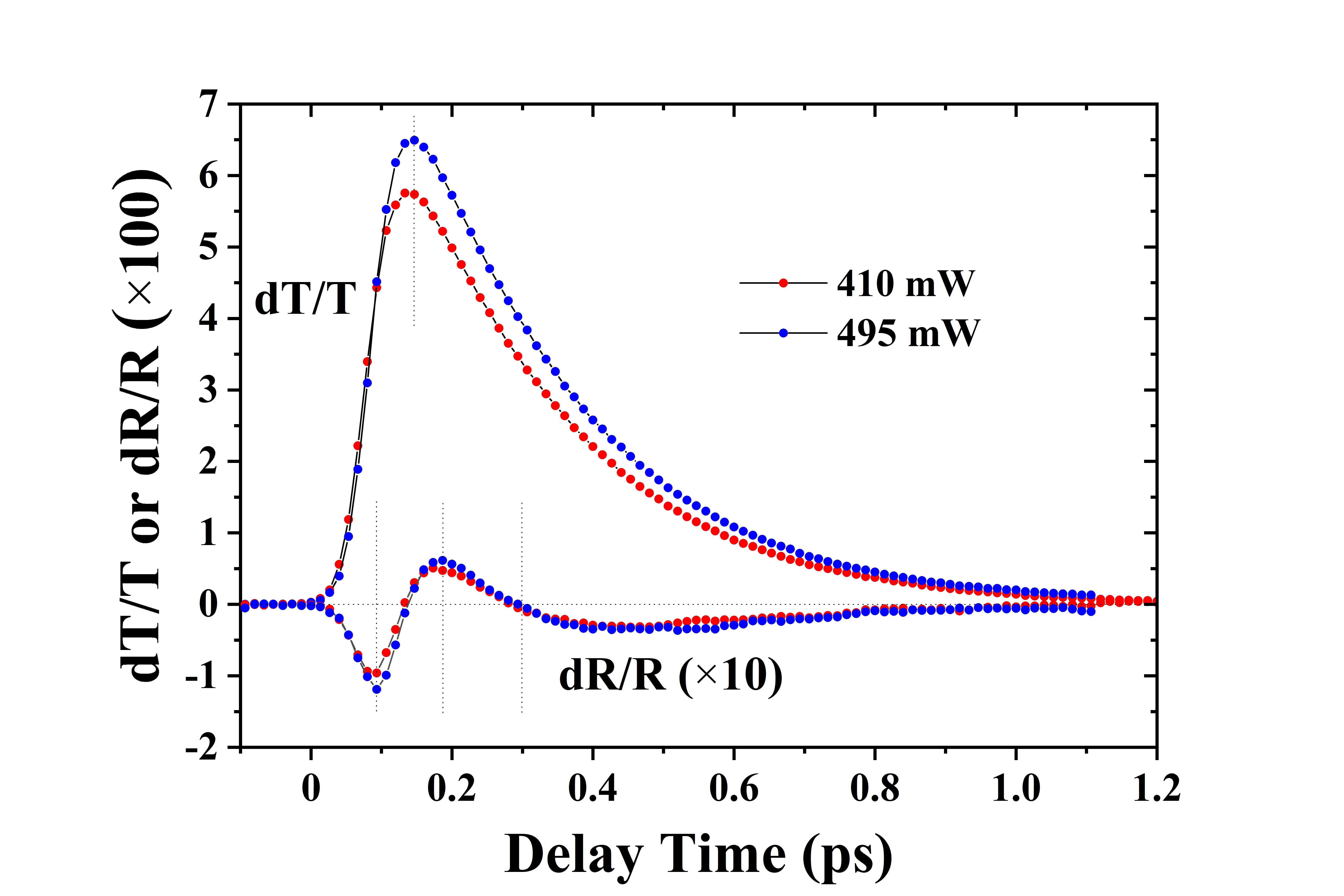}
\caption{The transition state of transient reflectivity for an excitation and probing wavelength of 775 nm. Here, the positive dR/R component is attributed to the Drude contribution caused by band filling of thermalized electrons, and the negative dR/R component is dominated by band gap renormalization.}
\label{775nm}
\end{center}
\end{figure}

Figure \ref{775nm} displays a critical transition state of the transient reflectivity at 775 nm. After intense excitation from the pump beam, the dR/R value first becomes negative; the value then rapidly changes to positive, and returns to negative again, consequently recovery to the initial state within 1 ps. Because the transient transmission and reflectivity were measured simultaneously, the origin of the delay time was set to the starting point. The results clearly show that the transient reflectivity responds more rapidly to an intense laser pulse, in contrast to the results for wavelengths above 775 nm, as shown in Fig. \ref{wave}. In contrast, the transient transimission remains positive, indicating the occurrence of optical bleaching. The obviously different spectral characteristics between transient transmission and reflectivity from simultaneous measurements suggest that these two processes originate from different physical mechanisms.

\section{Discussion}

\begin{figure}[h]
\begin{center}
\includegraphics[clip, width=8.0cm]{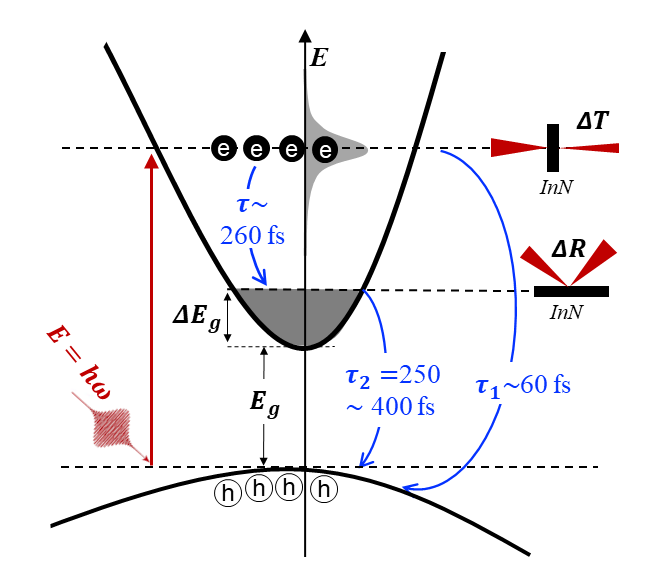}
\caption{Mechanism of optical bleaching and its recovery for the InN film, and measurement principles of transient transmission and reflectivity. $\tau$ is the time constant of relaxation from the excitation state to the CB edge, $\tau_1$ is the time constant for direct recombination between photoexcited electrons and holes, and $\tau_3$ is the time constant for band edge recombination. $E_g$ is the fundamental band gap of InN, and $\Delta E_g$ represents the band filling.}
\label{mechanism}
\end{center}
\end{figure}

Based on the above experimental observations, the mechanisms governing the transient transmission and reflectivity in InN thin films are schematically illustrated in Fig. \ref{mechanism}. After photoexcitation by a femtosecond pump pulse, electrons and holes are photogenerated at energy levels in the CB and valence band (VB), corresponding to vertical transitions in an $E$--$k$ band diagram at the pumping energy. When a large number of electrons are excited to the CB, optical bleaching occurs due to Pauli blocking,\cite{Iyer2017} which is a result of the occupation of photoexcited electrons at the probing energy level. In other words, the energy levels equal to the pumping photon energy are temporally occupied by photoexcited electrons, and allow the probe laser to be transmitted, thereby causing an increase in transient transmission. This physical picture differs from the previously reported band filling model,\cite{Pacebutas2006, Ahn2012} where partial filling of the CB leads to blocking of the lowest states and hence a widening of the optical band gap up to the probing energy level, which allows the probe to transmit rather than absorb. Our experimental results demonstrate that optical bleaching can occur as a consequence of the transient occupation of photoexcited electrons at the excitation levels.  

Our physical scheme suggests that the transient transmission spectra involve two decay processes: 1) the photoexcited electrons fall directly from the excitation states into holes in the VB (photoexcited electron--hole direct recombination), and 2) the photoexcited electrons cool down to the CB edge by various scattering mechanisms. Due to the screening of free carriers for the interaction between photoexcited electrons and holes, as discussed later, the direct recombination is negligible for the spectra in Fig. \ref{wave} (a) with a probing wavelength larger than 756 nm. The almost-exponential decay curves in Fig. \ref{802nm} (a) were fitted by an impulse response function of the form $A_{0}exp(-t/\tau)$, where $\tau$ is a time constant, and the time constant was extracted as $\sim$260 fs. In Fig. \ref{wave} (a), the transient transmission spectra match well at probing wavelengths larger than 756 nm. The numerical fit indicates that these spectra have similar time constants ($\sim$260 fs), which are associated with the time of relaxation from the excitation state to the CB edge, {\it i.e.}, the initial hot electron cooling rate.

On the other hand, the mechanism governing the change in transient reflectivity is complicated. The $s$-polarized reflectivity of the probe laser can be expressed by the Fresnel equation: 

\begin{equation}
 R_s  = \left| \frac{n_1 \cos\theta_i - n_2 \sqrt{1-(\frac{n_1}{n_2} \sin\theta_i )^2} }{n_1 \cos\theta_i + n_2 \sqrt{1-(\frac{n_1}{n_2} \sin\theta_i )^2}}  \right|, 
\end{equation}

where $n_2=\sqrt{\epsilon}$ is the complex refraction index of InN film, $\theta_i$=44.8$^\circ$ is the angle of incidence measured from normal direction in experiments, and $n_1$=1 is the refraction index of air. Based on the Fresnel equation, the intense laser irradiation is considered to induce a change in $n_2$, {\it i.e.}, a change in the complex dielectric constant $\epsilon$, which consequently gives rise to the transient reflectivity. Our numerical simulation shows that $R_s$ increases monotonically as a function of $\epsilon$.  

Generally, for most degenerated semiconductor materials, such as Al doped ZnO\cite{Jia2014} and Sn doped In$_2$O$_3$\cite{Jia2019},  the complex dielectric dispersion can be expressed as follows:

\begin{equation}
\epsilon(\omega)  =\epsilon_\infty + \epsilon_{g}(\omega) + \epsilon_{carrier}(\omega),  
\end{equation}

where $\epsilon_\infty$ is the high-frequency dielectric function of the medium, $\epsilon_{g}(\omega)$ describes the dielectric function due to the interband transition, and $\epsilon_{carrier}(\omega)$ represents the contribution of the carriers to the dielectric function. Without pump excitation, $\epsilon_{carrier}(\omega)$ is commonly determined by the Drude model, which describes the collective motion of free electrons or holes. Under the excitation state, $\epsilon_{carrier}(\omega)$ is considered from the following two contributions

\begin{equation}
\epsilon_{carrier}(\omega) =-\frac{\omega^2_{p,h}}{\omega^2+i\omega\Gamma_h} - \frac{\omega^2_{p,e}}{\omega^2+i\omega\Gamma_e}, 
\end{equation}

where the first term describes the contribution of photoexcited holes in the VB with the plasma frequency $\omega_{p,h}$ and the damping factor $\Gamma_h$, and the second term represents the contribution of thermalized electrons and unintentionally doped carriers in the CB with the plasma frequency $\omega_{p,e}$ and the damping factor $\Gamma_e$. Here, thermalized electrons denote photoexcited electrons that cool down to the CB edge. The plasma frequency $\omega_{p}$ is given by 

\begin{equation}
\omega_{p}=\sqrt{n_ce^2/\epsilon_\infty \epsilon_0 m^*},
\end{equation}

where $e$ is the elementary charge, $\epsilon_0$ is the permittivity of free space, $n_c$ is the electron or hole concentration, and $m^*$ is the effective mass of electron or hole. Here, the contribution of free electrons in the CB is considered to dominate the enhanced reflectivity amplitude in the measured NIR region (740 nm to 810 nm). This is because 1) the effective mass of the hole ($m_h^*$=1.65)\cite{Xu1993} is much larger than the effective mass of the electron ($m_e^*$=0.05)\cite{Fu2004}, and 2) the electron concentration (thermalized electrons + unintentionally doped carriers) in the CB is larger than the photoexcited hole concentration in the VB. By using $\epsilon_\infty$=6.7,\cite{Fu2004} $m_e^*$=0.05 and $n_e$=6.58$\times$10$^{19}$cm$^{-3}$, the plasmon energy was calculated to be $\hbar\omega_{p}$=0.49 eV for the unpumping sample. The damping constant $\Gamma_e$ is inversely proportional to the relaxation time of the electron, and can be calculated by $\Gamma_e$=e/$m^*\mu$, which is associated to a relaxation time $\tau_e$=8.17$\times$10$^{-16}$ s, indicating the existence of various scattering sources. 

Based on the above--mentioned physics, the rise in the reflectivity signal above 765 nm in Fig. \ref{wave} can be explained as follows. After excitation, their momentum and energy of the initial non-equilibrium hot electrons are rapidly randomized by various scattering processes; these electrons then cool down to the CB edge as thermalized electrons. The increase in carrier concentration in the CB gives rise to the enhancement in transient reflectivity. As the probe and pump move toward higher frequencies, the electrons, including thermalized electrons in the CB, become difficult to respond to the incident probing electric field, and their influence on the dielectric constant $\epsilon_{carrier}$ at high frequencies becomes weak. Therefore, the magnitude of the positive dR/R gradually decreases toward high probing frequencies, as shown for the dR/R curves ($\geq$785nm) in Fig. \ref{wave} (b). Below a certain wavelength, dR/R becomes negative.

The negative dR/R is attributed to band gap renormalization (BGR), which causes a decrease in the dielectric constant $\epsilon(\omega)$, {\it i.e.}, a decrease in the transient reflectivity. The BGR effect primarily involves 1) electron--hole interactions, and 2) electron--electron and hole--hole interactions. The electron--hole interaction has two terms: the screened Coulomb interaction and the bare Coulomb interaction.\cite{Benedict2002} The strong screened Coulomb interaction term is considered to dominate the BGR contribution, which is usually equivalent to the increase in photon energy. This interpretation is in reasonable agreement with our experimental observation that the absolute magnitude in negative dR/R in Fig. \ref{wave} (b) increases with the excitation energy increasing. On the other hand, the electron--electron interaction is due to the mutual exchange and Coulomb interaction between free electrons in the CB, and cause a downward shift of the CB.\cite{Feneberg2019, Hamberg1984} The VB is influenced in the opposite way due to the hole--hole interaction. Based on a weakly interacting electron-gas model, the band gap shrinkage due to the electron--electron interaction is approximately proportional to $n_c^{1/3}$.\cite{Wolff1962} In fact, the electron--electron interaction is the primary contributor to the BGR effect in transparent conductive materials with high carrier concentration.\cite{Hamberg1984}

Going back to the dR/R spectra in Fig. \ref{775nm}, the time evolution of the transient reflectivity can be described as follows. After the initial excitation, BGR occurs due to strong screening of the electron--hole Coulomb interaction and leads to a sharp drop in transient reflectivity. Next, rapid recovery occurs due to electron--hole recombination, and the screening of the Coulomb interaction becomes weak. Subsequently, as the photoexcited electrons thermalize to the CB edge by various scattering mechanisms, the contribution of thermalized electrons to the dielectric constant gives rise to the positive transient reflectivity. Consequently, as the thermalized electrons cool down further, the electron--electron interaction causes BGR, and hence, the dR/R becomes negative again. 

\begin{figure}[h]
\begin{center}
\includegraphics[clip, width=8.0cm]{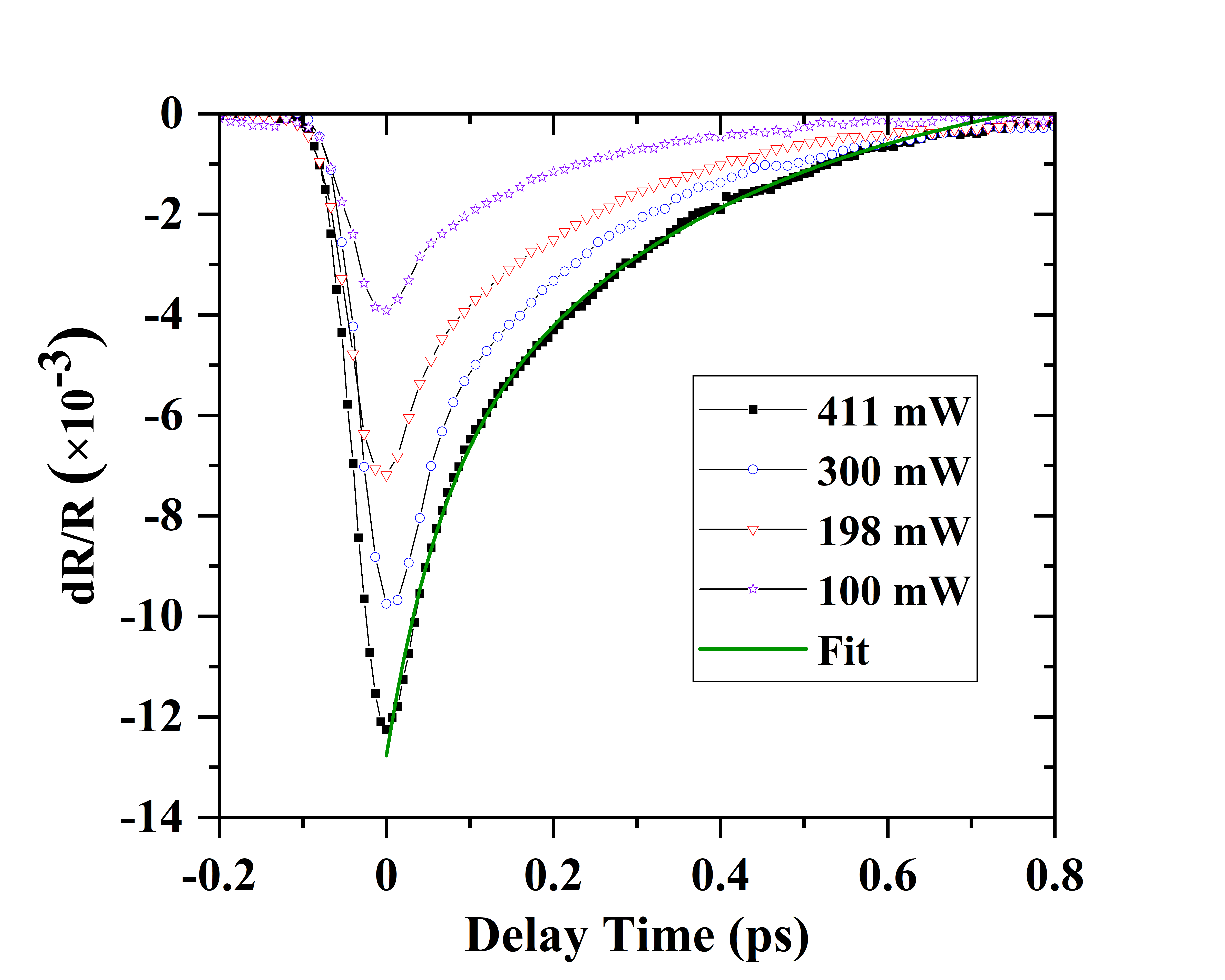}
\caption{Transient reflectivity spectra measured at an excitation and probing wavelength of 740 nm for different pump powers. The solid line shows the fitting curve by a biexponential fit $A_{1}(1-exp(-t/\tau_1))+A_{2}(1-exp(-t/\tau_2))$ with $\tau_1$=63 fs and $\tau_2$=377 fs as time constants.}
\label{740nm}
\end{center}
\end{figure}

Furthermore, we can gain insight into the BGR mechanisms by measuring the transient reflectivity at $\omega\tau_e >1$, becasue the carriers are difficult to follow the probing electric field and the interband transition $\epsilon_{g}$ dominates the dielectric function. Fig. \ref{740nm} shows a two-step recovery process, {\it i.e.}, the signal shows an initially rapid recovery followed by a slower recovery from the origin point, which corresponds to two BGR mechanisms at the excitation and probe wavelength of 740 nm ($\omega\tau_e=2.08$). A double exponential function was used to fit the recovery curves at different pumping powers, as shown in Table \ref{tao}. The extracted time constants $\tau_1$ almost remain close to 60 fs, and such ultrafast recovery is attributed to direct recombination between photoexcited electrons and holes due to the existence of a bare Coulomb interaction. The hot phonon effect should not be responsible for this ultrafast recovery after the initial excitation because the ultrafast Raman measurement indicates a 0.7 ps room-temperature LO phonon lifetime in InN,\cite{Pomeroy2005} which is much longer than our observed fast cooling time. Whereas, the time constant $\tau_2$ gradually increases from 250 to 400 fs as the pump power increases. This slow recovery is attributed to recombination between holes and electrons in the CB edge. Our results are consistent with previously reported electron cooling times of 400 fs.\cite{Sun2008, Su2010} Other studies reported longer decay times in the temporal development of the optical reflectivity difference, ranging from picoseconds to nanoseconds, which was recognized as carrier lifetime.\cite{Ascazubi2006, Shu2006}     

\begin{table}[h]
\caption{\label{tao} Recovery time constants were obtained from the best fit to the measured transient reflectivity spectra in Fig. \ref{740nm}, where a double exponential function $A_{1}(1-exp(-t/\tau_1))+A_{2}(1-exp(-t/\tau_2))$ with $\tau_1$ and $\tau_2$ as time constants was used.}
\begin{ruledtabular}
\begin{tabular}{rllc}
Pump power & $\tau_1$ (fs)  & $\tau_2$ (fs) &\\
\hline
411 mW     & 63$\pm$6    & 377$\pm$53   \\
300 mW     & 73$\pm$16  & 277$\pm$48   \\
198 mW     & 43$\pm$6    & 257$\pm$16   \\
100 mW     & 59$\pm$12  & 252$\pm$42   \\
\end{tabular} 
\end{ruledtabular}
\end{table}

The increase in $\tau_2$ with increasing the pump power implies that the electron--electron interaction suppresses the band edge recombination and hence mitigates the recovery process, which is consistent with Monte Carlo simulation results for the ultrafast relaxation of photoexcited carriers in GaAs.\cite{Osman1987} Recent first principle calculations also show that Auger recombination is suppressed by free--carrier screening at high carrier densities.\cite{McAllister2018} Experimentally, Chang et al. observed a $\tau$ $\propto$ $n^{-1/3}$ power law between photoexcited carrier density and average relaxation time on the order of 1 ps, indicating that carrier--carrier scattering is effectively restricted to only nearest--neighbor interactions as a result of the Coulomb screening in dielectric medium.\cite{Chang2007} Such experimental observations suggest that the momentum redistribution of photoexcited carriers is characterized by the scattering time for an electron to move out a sphere with average radius $r$ $\propto$ $n^{-1/3}$, which can be referred to the effective screening length. Following this scenario, our observed band edge recombination is speculated as Auger recombination, which needs the participation of another electron to release the excess energy. The decrease in $r$ leads to a reduced scattering frequency, {i.e.}, a reduced recombination possibility, and consequently mitigates the recovery process. 

It is worthy to mention that the response time of the carriers in the CB to screen the photoexcited hole depends on the excitation energy. The higher the energy of the photoemitted electron, the longer the response time, as shown in Fig. \ref{wave}. In this study, $\hbar\omega$=1.68 eV (740 nm) can be considered to approach the high frequency response limit, since the maximum transient transmission occurs simultaneously with  the minimum transient reflectivity, implying that the carriers cannot rapidly respond to screen the photoexcited holes. In this high--frequency limit, the Drude contribution is unimportant and the behavior acts like that for a dielectric. On the other hand, our experimental results also suggest that the dielectric function screens the electron--hole interaction in a very short response time within the scale of the Drude model, as proposed in the theoretical treatment in the random-phase approximation (RPA).\cite{Kubler} In addition, we observed a slight increase in the dR/R magnitude with increasing pump power in Fig. \ref{775nm}, suggesting that the screening of Coulomb interaction also associates with the excitation intensity. A high--intensity excitation causes an obvious BGR effect. 

\section{Conclusions}

We have performed the simultaneous measurements of ultrafast transient transmission and reflectivity on InN thin film under intense laser irradiation, and proposed a phenomenological physical scheme to explain optical bleaching and its recovery behavior. Based on the obvious difference between the transient transmission/reflectivity spectra characteristics, optical bleaching is attributed to Pauli blocking, which allows the probe light to be transmitted due to the occupation of photoexcited electrons at the probing energy level, rather than the traditional band filling model. Moreover, a systematical study on the transient reflectivity at different excitation/probing wavelengths revealed that the band filling and BGR effects dominate the spectral characteristics of the transient reflectivity, where the partial filling of thermalized electrons in the lowest states of CB leads to a positive dR/R, and the BGR effect caused by the screening of electron--hole Coulomb interaction and the electron--electron interaction gives rise to a negative dR/R. Based on these mechanisms, the time constants of photoexcited electron--hole recombination and band edge recombination are determined to be $\sim$60 fs and 250--400 fs, respectively. Our results also reveal that the electron--electron interactions suppress band edge recombination, and mitigates the recovery process. The simultaneous measurement of transient transmission/reflectivity opens a new way of studying the ultrafast dynamics of the electronic structure of semiconductors, and shows that the band structure is controllable by selecting the suitable excitation wavelength in semiconductors, which display variety of interesting physical phenomena.

\section{Acknowledgments}
J. Jia acknowledges the funding from JSPS KAKENHI Grant-in-Aid for Scientific Research (C) (Grant No. 20K05368), and this work is also a part of the outcome of research performed under a Waseda University Grant for Special Research Projects (Project number: 2020C--316).


\begin{thebibliography}{9}

\bibitem{Pacebutas2006} V. Pa{\v c}ebutas, G. Aleksejenko, A. Krotkus, J. W. Ager III, W. Walukiewicz, Hai Lu, and W. J. Schaff, Appl. Phys. Lett. {\bf 88}, 191109 (2006).

\bibitem{Ahn2012} H. Ahn, C. --C. Yu, P. Yu, J. Tang, Y. --L Hong, and S. Gwo, Optics Express {\bf 20}, 769 (2012).


\bibitem{Sun2008} S. --Z. Sun, Y. --C. Wen, S. --H. Guol, H. --M. Lee, S. Gwo, and C. --K. Sun, J. Appl. Phys. {\bf 103}, 123513 (2008).

\bibitem{Tsai2007} T.--R. Tsai, C.--F. Chang, and S. Gwo, Appl. Phys. Lett. {\bf 90}, 252111 (2007).

\bibitem{Ricardo2006} R. Asc{\'a}zubi, I. Wilke, S. Cho, H. Lu, and W. J. Schaff, Appl. Phys. Lett. {\bf 88}, 112111 (2006).

\bibitem{Ahn2009} H. Ahn, C.-H. Chang, Y.-P. Ku, and C.-L. Pan, J. Appl. Phys. {\bf 105}, 023707 (2009).

\bibitem{Wu2002} J. Wu, W. Walukiewicz, K. M. Yu, J. W. Ager III, E. E. Haller, Hai Lu, W. J. Schaff, Y. Saito, and Y. Nanishi, Appl. Phys. Lett. {\bf 80}, 3967 (2002).

\bibitem{Yu2005} K. M. Yu, Z. Liliental-Weber, W. Walukiewicz, S. X. Li, R. E. Jones, W. Shan, J. W. Ager III, E. E. Haller, H. Lu, and W. J. Schaff, Appl. Phys. Lett. {\bf 86}, 71910 (2005).

\bibitem{Iyer2017} V. Iyer, P. Ye, and X. Xu, 2D Materials {\bf 4}, 2 (2017).

\bibitem{Jia2014} J. Jia, N. Oka, M. Kusayanagi, S. Nakatomi, and Y. Shigesato, Appl. Phys. Express {\bf 7}, 105802 (2014).

\bibitem{Jia2019} J. Jia, A. Takaya, T. Yonezawa, K. Yamasaki, H. Nakazawa, and Y. Shigesato, J. Appl. Phys. {\bf 125}, 245303 (2019).

\bibitem{Xu1993} Y. -N Xu, and W. Y. Ching, Phys. Rev. B {\bf 48} 4335 (1993).

\bibitem{Fu2004} S. P. Fu, and Y. F. Chen, Appl. Phys. Lett. {\bf 85}, 1523 (2004).

\bibitem{Wolff1962} P. A. Wolff, Phys. Rev. {\bf126}, 405 (1962).

\bibitem{Benedict2002} L. X. Benedict, Phys. Rev. B {\bf 66}, 193105 (2002).

\bibitem{Feneberg2019} M. Feneberg, J. Nixdorf, C. Lidig, R. Goldhahn, Z. Galazka, O. Bierwagen, J. S. Speck, Phys. Rev. B {\bf 93}, 045203 (2016).

\bibitem{Hamberg1984} I. Hamberg, C. G. Granqvist, K. F. Berggren, B. E. Sernelius, and L. Engstrom, Phys. Rev. B {\bf 30}, 3240 (1984).

\bibitem{Osman1987} M. A. Osman, and D. K. Ferry, Phys. Rev. B {\bf 36}, 6018 (1987).

\bibitem{Pomeroy2005} J. W. Pomeroy, M. Kuball, H. Lu, W. J. Schaff, X. Wang, and A. Yoshikawa, Appl. Phys. Lett. {\bf 86}, 223501 (??2005)??.

\bibitem{Kubler} J. K{\" u}bler, Phys. Rev. {\bf 183}, 703 (1969).

\bibitem{Ascazubi2006} R. Asc{\' a}zubi, I. Wilke, S. Cho, H. Lu, and W. J. Schaff, Appl. Phys. Lett. {\bf 88}, 112111 (2006).

\bibitem{Shu2006} G. W. Shu, P. F. Wu, M. H. Lo, J. L. Shen, T. Y. Lin, H. J. Chang, Y. F. Chen, C. F. Shih, C. A. Chang, and N. C. Chen, Appl. Phys. Lett. {\bf 89}, 131913 (2006).

\bibitem{McAllister2018} A. McAllister, D. Bayerl, and E. Kioupakis, Appl. Phys. Lett. {\bf 112}, 251108 (2018).

\bibitem{Chang2007} Y. -Ming Chang, and S. Gwo, J. Appl. Phys. {\bf 102}, 083540 (2007).

\bibitem{Su2010} Y. -E Su, Y. -Chieh Wen, H. -Mao Lee, S. Gwo, C. -Kuang Sun, Appl. Phys. Lett. {\bf 96}, 052108 (2010).

\end{thebibliography}
\end{document}